\documentclass[A4paper,preprintnumbers,epsfig,aps,12pt]{revtex4}
\usepackage{amssymb}
\usepackage{graphicx,epsfig,chemarr}
\usepackage[usenames,dvipsnames]{color}
\usepackage{times}

\def\JPE#1{JPE: \color{red} {\tt #1}}

\def\be{\begin{equation}}
\def\ee{\end{equation}}
\def\ba{\begin{eqnarray}}
\def\ea{\end{eqnarray}}
\def\kbar{\bar{k}}

\def\local{{\text {local}}}
\def\nuc{{\text {nuc}}}

\def\JPE#1{\kern 30ptJPE:\ \color{red}{\tt #1}\color{black}}

\def\HALF{{\textstyle\frac{1}{2}}}
\def\eg{{\it{e.g.}}}
\def\ie{{\it{i.e.}}}
\def\ran{{\rm ran}}

\def\fs{{f_\ast}}
\def\fsran{{f_\ast^{\ran}}}
\def\Nlam{{N_\lambda}}
\def\ts{{t_\ast}}
\def\Ns{{N_\ast}}
\def\Rb{{\mathbf{R}}}
\def\rb{{\mathbf{r}}}
\def\dr {{\mathrm{d}}}

\begin{document}
\title{Remarks on Bootstrap Percolation in Metric Networks}
\author{T. Tlusty}
\affiliation{Physics of Complex Systems, Weizmann Institute of Science,  Rehovot, (Israel)}
\author{J.-P. Eckmann}
\affiliation{D\'epartement de Physique Th\'eorique et Section de
Math\'ematiques, Universit\'e de Gen\`eve, (Switzerland)}
\begin{abstract}We examine bootstrap percolation in
$d$-dimensional, directed metric graphs in the context of recent measurements
of firing dynamics in 2D neuronal cultures. There are two regimes, depending on
the graph size $N$. Large metric graphs are ignited by the occurrence of critical
nuclei, which initially occupy an infinitesimal fraction, $\fs \rightarrow 0$,
of the graph and then explode throughout a finite fraction. Smaller metric graphs are
effectively random in the sense that their ignition requires the initial
ignition of a finite, unlocalized fraction of the graph, $\fs > 0$. The
crossover between the two regimes is at a size $\Ns$ which scales
exponentially with the connectivity range  $\lambda$ like $\Ns \sim
\exp\lambda^d$. The neuronal cultures are finite metric graphs of size
 $N \simeq 10^5-10^6$, which, for the parameters of the experiment,
is effectively random since $N \ll \Ns$. This
explains the seeming contradiction in the observed finite $\fs$ in these
cultures.  Finally, we discuss the dynamics of the firing front.
\end{abstract}
\maketitle

\subsection{Introduction}

Percolation in directed graphs deals with propagation of firing of
nodes whereby each node which was ``on'' in turn ignites any node that
can be reached by following the directed links.
Bootstrap percolation (BP) generalizes conventional percolation by applying
a stricter definition of ignition, a site only ignites when $m>1$
sites pointing to it were ignited earlier
\cite{Chalupa-1979,Kogut-1981,Adler-1991,Bollobas-1984}.
This generalization proved to be useful in many contexts (see \cite{Adler-1991,Seidman-1983} and references therein).
Here, we have in mind questions in connection with recent measurements of 2D
neuronal cultures (NC) \cite{Breskin-2006,Eckmann-2007,Soriano-2008,Cohen-2009}.

In these NC experiments, one grows a 2D culture of $N\sim 10^5-10^6$ neurons.
After the culture develops connections and forms a network,
 one externally ignites a variable number of neurons by varying a
``firing threshold'' in the culture, and observes how many neurons
eventually will have fired. The experiments show that, as
the fraction of initially (\ie, externally) excited neurons exceeds a certain
\textit{non-zero} critical value, $\fs>0$, the activity of the network jumps abruptly to a
basically complete ignition of the network within a very short time,
while below this threshold, the firing essentially does not spread.

Thus, the experiment exhibits a sharp transition in the number of
excited neurons
\cite{Cohen-2009} when a certain initial firing threshold (fraction) is
exceeded. A simple theoretical model was able to capture this phenomenon by
describing the network in terms of a process very similar to BP in a random
graph. There are some paradoxes in this result which
the current paper will resolve: The random graph model lacks an important
ingredient of actual neuronal networks in assuming that all pairs of neurons
have equal probability to be connected. This is in contrast to the
experimental reality that the connection probability of close-by neurons is
much higher than that of distant ones \cite{Soriano-2008}.

Taking these distances into account, we are in the realm of \textit{embedded}
networks, where the neurons have a fixed position in physical space and, as
a consequence, the neural connections have a certain distance. This leads to
an apparent paradox, because it is known---and will be discussed
below---that in embedded graphs that have a metric, such as lattices, it is
enough to externally excite an infinitesimal fraction of the neurons, $\fs \rightarrow 0$,
to activate a finite, non-zero fraction of the network. This would seem to
contradict the experimental findings of \cite{Cohen-2009} with their
finite threshold. Or, asking
differently: (i) Why does the random graph picture of \cite{Cohen-2009}
describe so successfully the measurements of a 2D NC although it completely
neglects notions of space and vicinity? (ii) More generally, when can
real-space networks be described as effectively random
networks, and under which conditions does one have to take space into
consideration?

As we explain below, the puzzle is resolved because there is a basic
difference in the manner in which random and metric graphs are ignited. In
the latter, it suffices to initially turn on localized excitation \textit{%
nuclei}, which are then able to spread an
excitation front throughout the spatially extended network. In contrast, by
their definition, there are no such nuclei in the random graphs,
which lack the notion of locality, and one has to excite a \textit{finite fraction}
of the neurons to keep the ignition going. Still, as we shall show,
the experimental network---which is obviously an example of a metric
graph---is effectively random, since its finite size makes the
occurrence of excitation nuclei very improbable (an answer to question (i)).
More generally, we see that one can change the effective behavior of a network by
changing its finite size $N$ or by changing the finite range of connectivity $\lambda$
(question (ii)). For infinite graphs, what matters is the manner in which these
two quantities approach infinity.
The crossover between effectively random and metric behaviors is determined by a size $\Ns$,
which explodes exponentially with the connectivity range, $\Ns \sim \exp \lambda^d$,
where $d$ is the dimensionality.
We also discuss the dynamics of the propagating firing cluster in effectively metric and random graphs.

\subsection{The basic setting}

We consider ``neural networks'', that is, directed graphs with $N$
``neurons'', by which we mean nodes connected by directed links.
We assume
that any ordered pair $i\ne j$ of nodes can only be connected by one link.
The connections are described by the adjacency matrix $A$, with $A_{ij}=1$
if there is a directed link from $j$ to $i$, and $A_{ij}=0$ otherwise.

The neurons fire, and once they fire, in the model we consider here, they
stay ``on'' forever. A neuron is supposed to be on at the beginning of time
with some probability $f$ and will be on at time $t+1$ if at time $t$ it was
on, or if at least $m$ of its upstream (incoming) nodes were on at time $t$ (%
$m$ may be termed ``minimal influx"). This is described by the evolution
equation
\be
s_{i}(t+1)=s_{i}(t)+\left( 1-s_{i}(t)\right) \theta
\left(
\sum_{j}A_{ij}s_{j}-m\right) ,  \label{dynamics}
\ee
where $s_{i}(t)$ describes the state of the neuron at time $t$:
It is on if $s_{i}(t)=1$ and off if $s_{i}(t)=0$. Finally, $\theta $ is the
step function.

This setting is practically equivalent to BP (which is also known as ``$k$-core percolation").
In a typical BP scenario, the nodes of a graph or lattice, usually undirected ones, are randomly
populated and those nodes having less than $k$ edges are
``pruned" \cite{Adler-1991}. After iterative pruning there remains a connected
``$k$-core", which may vanish depending on the system
parameters. It is straightforward to see that the analogue of pruning is
the propagation of the firing cluster throughout the NC; the nodes that are
pruned in each iteration of BP are equivalent to the newly excited neurons
in each time step of the NC dynamics. In fact, if the network is $d$%
-regular, then a NC with a minimal influx of $m$ firing inputs can be mapped
to a \textit{directed} BP process with $k=d-m$. In the context of NC, one may term
the dynamics of firing propagation ``quorum percolation" \cite{Cohen-2009}
since the ignition of a certain neuron requires a ``quorum" of $m$ firing inputs.
It also hints for the potential uses of these models to describe the spread of diseases, rumors and opinions.

Some facts are obvious from the definition of the model: The dynamics is
monotonic, since a firing neuron can never turn off and therefore $%
s_{i}(t+1)\geq s_{i}(t) $. Therefore, for any initial condition ${\underline
s}(0)\equiv\{s_{i}(0)\}_{i=1}^N$ and any $j$ one has that $\lim_{t\to\infty}
s_j(t) \equiv s_j(\infty )$ exists. Furthermore, if $N$ is finite, then the
system (\ref{dynamics}) converges to a steady state (in finite time).

The dynamics is conveniently characterized by the initial firing
concentration, $f$ defined by $f_N({\underline s},0)$, where, for any $t$,
\[
f_N({\underline s},t)=N^{-1}\sum_{j=1}^N s_j(t).
\]
It has been observed \cite{Cohen-2009} that in directed
random graphs, with fixed degree distribution, there is a value $\fs > 0$ (depending on
this degree distribution) such that for large enough $N$ and for
\[
f_N({\underline s},0)>\fs
\]
there is ``substantial firing'', characterized by
\[
\lim_{N\to\infty } f_N({\underline s},\infty ) > f_N({\underline s},0)~,
\]
for ``many'' random graphs and for ``many'' initial distributions ${%
\underline s}$ (in a suitable measure theoretic sense, for example with
probability 1 with respect to the uniform measure). For $m=1$ all this
is in
the realm of percolation theory.

\subsubsection{Random graphs}

The study of this problem for directed random graphs can be found in
\cite{Cohen-2009}.
 Based on an ensemble average of (\ref{dynamics}) it yields a
self-consistency equation for the fraction $\Phi \equiv \lim_{t\rightarrow
\infty }f_{N}(\cdot ,t)$ of typically lit neurons at infinite time (in
practice, when stationarity has been reached, no later than $t= N$):
\be
\Phi =f+(1-f){\text{Prob}\ }(\#\text{firing inputs}\geq m)=f+(1-f)\Psi
(m,\Phi ),  \label{Maria}
\ee%
where the collectivity function $\Psi (m,p)$ accounts for the combinatorics of
choosing at least $m$ firing inputs of a randomly chosen node, when each
node fires with a probability $\Phi $. Assuming the probability for a node
to be of in-degree $k$ is $p_{k}$, the combinatorial expression for $\Psi $ is

\be
\Psi (m,\Phi )=\sum_{k=m}^{\infty }
p_{k}\sum_{\ell=m}^{k}\dbinom{k}{\ell}\Phi^{\ell}\ (1-\Phi )^{k-\ell}.
\label{Psi}
\ee
For similar treatments of undirected lattice and random graphs see
\cite{Chalupa-1979,Dorogovtsev-2006,Balogh-2007}.
When the average in-degree $\kbar=\sum_{k}{k p_{k}}$
and the minimal influx $m$ are large, $\kbar,m\gg 1$,
one can neglect the variance in the number of firing inputs
and the number of firing inputs into a node of degree $k$ is $\sim k\Phi $.
As a result, $\Psi (m,\Phi )$ can be approximated by its mean-field expression
$\Psi (m,\Phi )\simeq {\text{Prob}\ }(k\Phi\geq m)
\simeq \int_{k=m/{\Phi}}^{\infty }p_{k}$, see \eg, \cite{Sergi-2005}.
In the case of a regular graph, with in-degree $k=\kbar$ at all nodes, $%
p_{k}=\delta _{k,\bar k}$, one finds $\Psi(m,\Phi) \simeq \theta (\Phi-m/\kbar )$.

Equation (\ref{Maria}) can be rewritten as a function from $\Phi$ to $f$,
$f=F(m,\Phi )=(\Phi -\Psi)/(1-\Psi)$,
or, using the inverse function, as
$\Phi(m,f) =F^{-1}(m,f)$.
The solution of the latter equation jumps from
$\Phi \simeq f$ to $\Phi \simeq 1$
when $\partial_\Phi F=0$ (\ie, $F^{-1}$ is multi-valued),
or equivalently when
$\partial_\Phi \Psi|_{\Phi _\ast }=(1-\Psi |_{\Phi _\ast })/(1-\Phi _\ast )$.
It follows from the definition of $\Psi $ that there exists such critical
$\Phi_\ast$ and $\fs=F(m,\Phi _\ast )>0$.
In the simple case of the regular graph, neglecting fluctuations
in the mean field approximation implies that every node has exactly $\kbar\Phi$
firing inputs and, as a result, $\fs \simeq m/\kbar$.
This mean field result approximates $\fs$ well
when the in-degree has a large average $\kbar$ and relatively small variance.

\subsubsection{Metric graphs}

Our aim here is to consider not only connectivity but also metric, mimicking
in a poor man's way the idea that neurons have axons of finite length and
are located at some position in $\Rb^d$ (with $d=2$ or $d=3$,
depending on the experiment).

Things change rather drastically, when the notion of metric and proximity is
added to the game, for example by putting the neurons on a lattice. In this
case, we can show that $\fs=0$ no matter what $m$ is as long as the
coordination number in the lattice exceeds $m$.
\begin{figure}[ht]
  \begin{center}
 \includegraphics[scale=0.65]{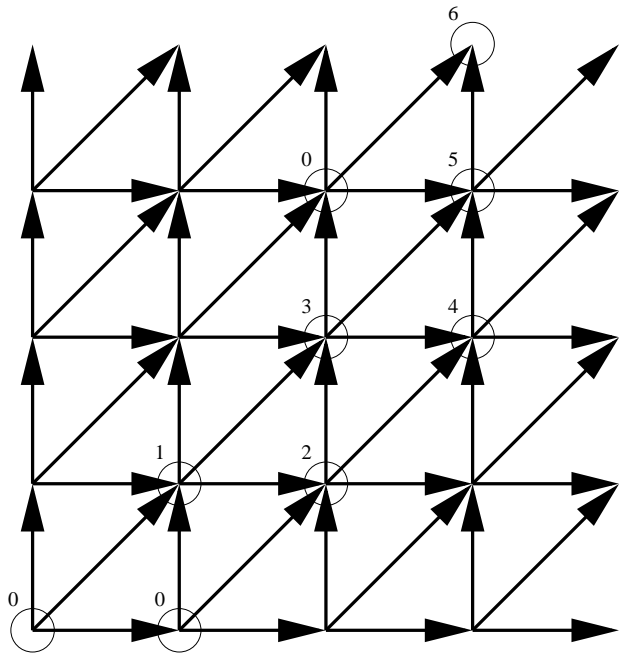}

  \caption{A simple example of a directed graph  with $k=3$ and $m=2$,
    where 2 initially ignited sites generate an infinity of ignited sites along a
    diagonal strip. The times of ignition are indicated by the numbers next to the circle.
    Also note that the ignited strip gets thicker (from 2 to 3 diagonals)
    when the site numbered 5 is close to another site which was initially ignited.}
\label{f:strip}
  \end{center}
\end{figure}

A simple example is the triangular lattice of Fig.~\ref{f:strip} where, for each node, the $3$ edges
pointing to the North, East and NE are outputs and the $3$ edges
coming from S, SW and W are inputs. For $m=2$, it suffices that somewhere in
the
lattice two neighboring nodes fire to ignite a finite fraction of the
lattice (\ie, $\fs=0$). First, a diagonal stripe propagates through
the lattice. When this diagonal meets a firing node at a neighboring
diagonal it ``activates" this diagonal which is now included in the
propagating stripe. Eventually, the stripe thickens and might cover the whole
lattice \footnote{
Strictly  speaking, the propagation of the fire through the graph is not
connected to the metric properties, but only to the  topology. However,
here, and in the other examples, the topology is  most simply described by
mapping the nodes of the graph in  $\Rb^d$ and describing
connectivity as a function of distance.}.

Note that what really matters here is the appearance of a \textbf{firing
nucleus}. Namely, if {\emph{somewhere}} in the graph 2 neighbors are lit up
at time 0, then a sizeable part (or all) of the graph will burn. Similarly,
when $m$ inputs are needed, a nucleus of size (proportional to) $m$ will
suffice to ignite a large burning within the graph.

Our focus is on the transition between the two BP scenarios we have
described so far: The random non-metric graph and a graph which is embedded
in a metric space. Apparently, the question in the metric graph is that of
the probability of a nucleus being lit up at time 0. In the next section, we
propose a model which comes relatively close to the experimental setup as
described in \cite{Breskin-2006,Cohen-2009}.

\subsection{ Networks in space}

Consider the following model of a spatial network: A large number $N$
of nodes
are distributed randomly in $\Rb^d$ with a number density $n=N/V$.
In other words, there are, on average, $n$ nodes per unit volume. We now
give ourselves a \textquotedblleft vicinity function\textquotedblright\ $g:%
\mathbf {R}^{+}\rightarrow \mathbf {R}^{+}$, with the property:
\be
n\int \dr^{d}x\,g(|x|)=\kbar~.
\ee
A directed edge connects any two nodes with a probability (correlation) $g(r)
$ that depends on the distance $r=\left\vert x_{i}-x_{j}\right\vert $.
Typical examples of functions $g$ are (normalized versions of) $g(r)\sim
\exp (-r/\lambda )$, $g(r)\sim \exp (-\left( r/\lambda \right) ^{2})$ or a
scale-free type $g(r)\sim r^{-\alpha }$. For the scale-free graph we define
an effective length scale $\lambda _{e}$ as the radius of the sphere drawn
around a node that contains "most" of the nodes that are connected to
central one, \eg, 99\% of them.

\subsubsection{Infinite graphs}

\label{disturbed}

There is an essential difference between finite, but large, graphs and infinite graphs.
We first deal with the infinite graph. Since the graph is infinite, one can
keep $f$ arbitrarily close to 0, and still have, somewhere in this infinite
graph, any desired firing nucleus. Of course, the probability to find, in
any finite region, a big nucleus is exponentially small in the number of
nodes which are supposed to be lit up in a cluster at time 0.

Still, in infinite volume, and for any $f>0$, we can assume that there is a
nucleus much bigger than the scale $\lambda $ of $g(r)$, resp.~much bigger
than $\lambda_e$ when $g$ is like a power law, and that inside of this nucleus the
firing fraction at time 0 is a certain $f_\local$ (which one can
choose as close to 1 as one wishes).

Consider now a non-lit node which is
close to the boundary of such a nucleus (and assume for simplicity that the
nucleus contains a large sphere of lit nodes). There will also be some
``hair'' lit up, but we neglect this, since it only will make propagation
stronger.
On average, half (if the sphere is big enough) of the inputs of the given
node come from the sphere. It follows that to fire this node it is enough to
have $f_\local \kbar/2 \geq m$ (and one can take any $f_\local$, \eg, $1$
since we are considering an infinite graph). As a consequence,
the firing sphere will explode throughout a large part of the graph.

There are some corrections to this simple mean-field condition, which take
into account the combinatorics of choosing the $m$ (or more) firing inputs
from the sphere. However, these corrections are negligible if $\kbar,m\gg 1$.
The corrections amount to solving (\ref{Maria}) with $k_\text{eff}=k/2$
and $f_\local$. In summary, to fire a spatial network all that
is needed is a large enough nucleus ($\gtrsim \lambda $), which will be
found in an infinite graph (but not in experimentally available NC graphs) even
at infinitesimal $f$.

Thus, we have the following dichotomy: either $\lambda $ is finite and then $%
\fs=0$ as we argued above, or at $\lambda \rightarrow \infty $ we
return to the random graph with its finite $\fs>0$ because then any
two nodes are connected with the same probability. The argument above
applies also to a scale-free network when $\lambda $ is replaced by the
effective scale $\lambda _e.$

\subsubsection{Finite graphs and comparison with experimental data}

The ignition of a metric graph requires the appearance of a large enough nucleus.
It is therefore reasonable to define, in a statistical manner, that a graph is ignited
when the probability to find at least one such nucleus, $p_\nuc$, becomes non-negligible (say 50\%).
Ignition occurs when the initial firing, $f$, exceeds a critical value $\fs$.
In metric graphs, as we show below, the ignition point is a function
of the graph size $N$, the connectivity range $\lambda$, the average in-degree $\kbar$
and the minimal influx, $m$.
In the following, we heuristically estimate $\fs(N,\lambda,\kbar,m)$.
We show that there is a graph size, $\Ns$,
where the graph crosses over from an effectively random regime,
$\fs \simeq \fsran>0$ (the superscript ``ran" denotes the value for a random graph),
to the effectively metric regime, $\fs \rightarrow 0$.

We consider a metric graph of $N$ nodes and density $n = N/V$. The nodes
connect according to a vicinity function $g$ with a scale $\lambda$. At $t=0$ a
randomly chosen fraction $f$ of the nodes is ignited. For our purposes, we can
assume that the critical nucleus is approximately a sphere of radius $\lambda$,
containing $\Nlam \sim n \lambda^d$ nodes.

To ignite the graph, the local initial firing fraction $f_\local$
inside the nucleus must exceed $\fsran\simeq m/\kbar$ of the
corresponding random graph. We therefore look for the circumstances
when it is likely to find at least one such lit nucleus. Typically,
$\Nlam \gg 1$ so that the number $L_\lambda $ of lit nodes inside the
sphere distributes normally with mean $f \Nlam$ and variance $\sigma^2=f(1-f)\Nlam$.
The probability $s_\lambda $ that the sphere of radius $\lambda $
becomes critical is therefore
\be
s_\lambda = \Pr (L_\lambda  \ge \fsran \Nlam)
                 = \frac{1}{\sigma\sqrt{2\pi} }
                     \int_{\fsran \Nlam}^{\infty}\dr L_\lambda
                          \exp\left[ - \frac{(L_\lambda  - f\Nlam)^2 }{2\sigma ^2} \right]
                    \simeq \frac{1}{z\sqrt{2\pi}}  \exp\left(-\HALF z^2 \right),
\label{slambda}
\ee
where
\be \label{e:6}
z =\frac{\fsran \Nlam -fN_\lambda}{\sigma}
  =\frac{\fsran-f}{\sqrt{f(1-f)}}\Nlam^{1/2}
\ee
is the normalized deviation
of the required number of lit nodes $\fsran \Nlam$ from its mean $f \Nlam$.
To obtain the $\simeq$ in (\ref{slambda}) we have simply
replaced the integral by its lower boundary value (which is a good
approximation for the large $z$ we consider).

\begin{figure}[ht]
  \begin{center}
 \includegraphics[scale=0.65]{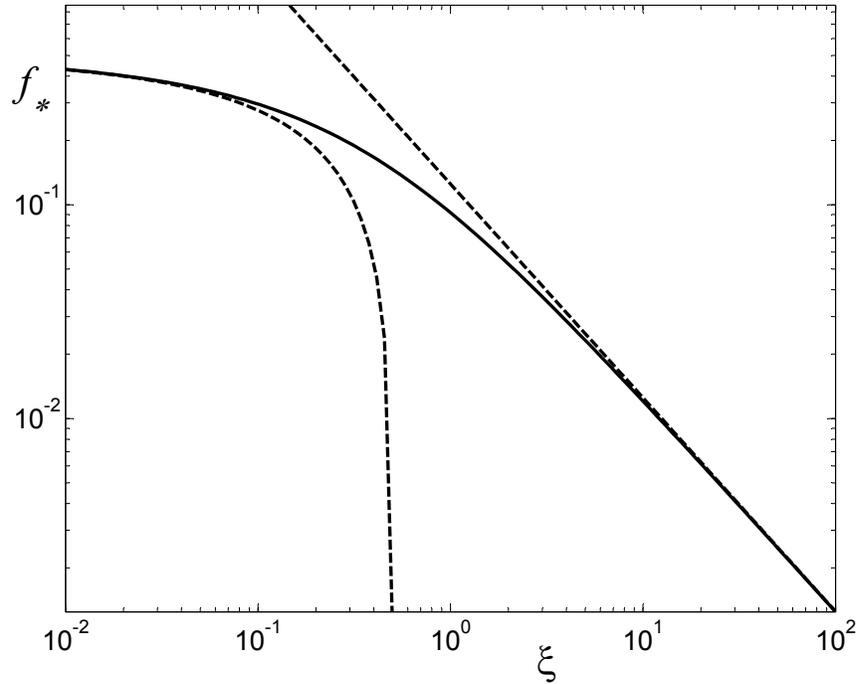}

  \caption{The minimal firing fraction $\fs$ is plotted as a function of the parameter
  $\xi = \log N/\Nlam$. The solid line is the solution of (\ref{e:99}),
  $\fs = (2\xi+1)^{-1}
  \left( \fsran + \xi - \sqrt{\xi(\xi+2\fsran(1-\fsran))}\right)$,
  whereas the dashed lines are the asymptotic solutions
  $\fs \sim \xi^{-1}$, in the metric regime (\ref{e:metric}), and
  $\fsran - \fs \sim \xi^{1/2}$ in the random regime (\ref{e:random}).  }
\label{f:xi}
  \end{center}
\end{figure}

The nucleation probability $p_{\nuc}$ becomes relevant when the expected
number of nuclei, $N s_\lambda$, exceeds one. This defines a crossover
condition,  $s_\lambda  \simeq N^{-1}$, which, by inverting (\ref{slambda}),
leads, for large $N$, to
\be
z_\ast \simeq (2\log N)^{1/2}~.
\label{zast}
\ee
This relation determines the fraction $\fs$ required to ignite a graph of size
$N$, vicinity scale $\lambda$ and when requiring a value
$\fsran$ for a random graph with the same connectivity (\ie,
degree distribution $p_k$) but no particular metric properties (Fig. \ref{f:xi}).
We find from (\ref{e:6}-\ref{zast}) that the fraction
$\fs$ at
the ignition transition is given by
\be \label{e:99}
\frac{(\fsran  - \fs)^2}{\fs (1-\fs)} \Nlam \simeq 2\log N~.
\ee

Solving (\ref{e:99}) for $\fs$, one finds that if $\fs$ is very small
which is the case for \emph{metric} graphs, then to ignite this metric graph
of size $N$ one would need to externally ignite a fraction
\be \label{e:metric}
\fs \simeq  \HALF (\fsran)^2 \frac{\Nlam}{\log N}~.
\ee
 Indeed, this fraction vanishes for infinite graphs, but only logarithmically.

At the other extreme, of effectively random graphs, $\fs$ is somewhat
larger, from (\ref{e:99}) we obtain
\be \label{e:random}
\fs = \fsran - \left[2 \fsran (1-\fsran)\right]^{1/2}
        \left( \frac {\log N}{\Nlam} \right)^{1/2}~.
\ee
In this case, as announced earlier, $\fs$
increases with $\Nlam \sim n \lambda^d$ and decreases logarithmically with
the size of the network, $N$.
A reasonable definition of the crossover between effectively random and
effectively metric graphs is the size $\Ns$ at which the required $\fs$ is
significantly reduced relative to the random graph value, $\fsran$, say
$\fs=\alpha \fsran$ with, \eg, $\alpha = 1/2$ (Fig. \ref{f:xi}).
From (\ref{e:99}), we find that this occurs at
\be \label{Ns}
\Ns \simeq \exp \left[
            \frac{(1-\alpha)^2\fsran}{2\alpha(1-\alpha\fsran)}\Nlam\right]
            \simeq \exp \left(
            \frac{\fsran}{4 -2\fsran}\Nlam\right) ~.
\ee

It is clear from the discussion that the relevant
dimensionless parameter in the system is $\xi \equiv \log N/ \Nlam \sim
\log N / \lambda^d$. The random graph regime is $\xi \ll 1$ whereas graphs are
effectively metric in the regime $\xi \gg 1$. The critical fraction $\fs$
changes its scaling from $ \fsran - \fs \sim \xi^{1/2}$ in the random
graph regime to $\fs \sim \xi^{-1}$ in the metric regime.

In the  NC experiment \cite{Breskin-2006,Eckmann-2007,Soriano-2008,Cohen-2009},
the 2D density of the neurons ranges between $n =150-1,000$  neurons/mm$^2$
and their average connectivity is $\kbar = 60 - 150$ inputs/neuron.
The connectivity scale is approximately $\lambda = 1 - 2$ mm \cite{Soriano-2008}
and the minimal nucleus size is therefore about $\Nlam = \pi n \lambda^2= 500 -10,000$ neurons.
The minimal influx in these experiments is $m\sim15$ before a drug
that weakens the synapses is added, which increases the minimal inputs to
$m\sim80$. These values correspond to minimal ignition fractions of $\fsran = 0.1-0.25$
for the drug free NC and $\fsran = 0.5-1$ after the drug was added.

By substitution of the experimental values in the formula for $\Ns$ (\ref{Ns}),
we find that the range of crossover size for drug free NC is  $\Ns =3\cdot10^5 - 10^{300}$.
Note that the lower regime is obtained for the extremal conditions of a very dilute culture
$n =150$ with low connectivity range,  $\lambda = 1$ mm,
which is also highly connected, $\kbar = 150$.
In fact, this combination of parameters is not feasible, since dilute networks tend to be less connected,
and the crossover range $\Ns$ is well above the experimental size of $N \sim 10^5-10^6$ neurons.
When drug is added, $\Ns$ explodes to  $\Ns \sim 10^{70} - 10^{2000}$.
These ridiculously large numbers --
the number of atoms in the universe is somewhere around $10^{80}$
whereas the number of neurons in the human brain is $\sim 10^{11}$--
imply that the NC can safely be regarded as a random graph
despite the fact that it is actually a metric graph embedded in space.
In other words, it is highly improbable to find an igniting nucleus and the NC
ignites ``homogeneously" at $\fsran$.
Therefore, geometry does not seem to come into play in those experiments,
only topology.

Interestingly, the situation in physical models of BP is quite the opposite.
In typical situations, one would have $\lambda \sim 1-2$ lattice constants or
$\Nlam \sim 3-10$
(for example, in the triangular lattice from Fig. \ref{f:strip}, $\Nlam =3$).
This implies moderate crossover size $\Ns \sim 10-10^5$.
Thus, unlike NC experiments, even relatively small graphs, of size
easily accessible to experiments and simulations, are well within the metric regime, where
$\fs \rightarrow 0$ (see \eg, \cite{Adler-1991}).
It also renders NC as a rare example for naturally occurring, {\it large} (effectively) random graphs.

\subsection{Another mechanism for transition in infinite graphs}

As we said, as long as the graph is infinite, there is a trivial phase
transition in $\fs$: either the graph has some scale, $\lambda < \infty$,
and $\fs=0$, or it does not and $\fs>0$. The transition between these
two behaviors appears at $\lambda =
\infty$.
As a side remark we note that one can shift the transition
to a finite $\lambda$ by the
following procedure: Divide the edges into two sub-populations: one that
spreads uniformly without any distance dependence and one that forms
according to $g(r)$. The probability that an edge connects $i$ and $j$
is
then
$%
g(|r_{i}-r_{j}|)+k_{\infty }/N$, where $k_{\infty }\leq \kbar$ is the
number of \textquotedblleft random" edges per node. This mechanism enables a
smoother transition. For any $\fsran$, there is a transition
connectivity, $k_{\infty }^{c}$
given by $1-k_{\infty }^{c}/\kbar=\fs$, where the system changes its
behavior.
When $k_{\infty }>k_{\infty}^{c}$ no nucleus is enough because it is now
impossible to surpass $f_{\ast
}^{\ran}$ locally and the graph needs to compensate by a uniform
$\fs=(k_{\infty
}-k_{\infty }^{c})/\kbar>0$. The parameter $k_\infty$ may be varied by
varying the connectivity range, $\lambda$.

\subsection{Dynamics}

Having dealt with the total number of lit nodes at ``infinite'' time, we
next discuss the dynamics of how this state is reached. One may think of a
dynamical system for the space-dependent ensemble average $\Phi(\rb%
,t) = \left\langle s_i(t) \right\rangle $ in the vicinity of $\rb$.

The average is for example over a ball of radius $\lambda $:
\[
\left\langle s_i(t) \right\rangle \equiv \frac{\sum_{j:|r_j-\mathbf{%
r|<\lambda }} s_j(t)}{\sum_{j:|r_j-\mathbf{r|<\lambda }} 1}
\]

If $\Phi(\rb,t)$ is smooth enough we can write a dynamical equation
\be
\Phi(\rb,t+1) = \Phi(\rb,t)+\left( 1 - \Phi(\rb%
,t)\right) \theta \left( n \int \dr\mathbf{r^\prime} \,g(\left\vert
\mathbf{r^\prime}-\rb\right\vert )\Phi(\mathbf{r^\prime},t)-m\right) ,
\ee
where $g$ is the vicinity function with its scale $\lambda $ as before.

The time continuous version is
simply
\be
\frac{\partial \Phi(\rb,t)}{\partial t}=\left( 1-\Phi(\rb%
,t)\right) \theta \left(n \int \dr\mathbf{r^{\prime \prime }}\,
g(\left\vert \mathbf{r^{\prime \prime }} \right\vert )\Phi(\mathbf{%
r+r^{\prime \prime }},t)-m\right) ,
\ee
with the change of variables $\mathbf{r^{\prime \prime }=r^{\prime }-r}$. If
$\Phi(r,t)$ is smooth over a scale $\sim\lambda $, then we can approximate
\[
\Phi(\mathbf{r^{\prime }},t)\sim \Phi(\rb,t)+\mathbf{r^{\prime \prime
}}\cdot \nabla \Phi(\rb,t)  + {\textstyle\frac{1}{2}} \mathbf{%
r^{\prime \prime }}\cdot \nabla^2 \Phi(\mathbf{\ r},t)\cdot \mathbf{%
r^{\prime \prime }}
\]
One finally gets

\be
\frac{\partial \Phi(\rb,t)}{\partial t} = \left( 1 -
\Phi(\rb,t)\right) \theta
\left( \kbar \Phi(\rb,t) + r_0^2 \nabla ^{2}\Phi(\rb,t) - m
\right),
\ee
where $r_0^2 = \frac{1}{2d} \int \dr \mathbf{x}\left\vert \mathbf{x}%
\right\vert ^{2}g(\left\vert \mathbf{x}\right\vert )$.
This intuitive result basically says that
$\Phi(\rb,t)$ increases at a rate that is a product of the
probability that $\rb$ was not already ignited and a very steep function of the
average number of firing neurons in its vicinity. This averaging is performed
by taking the Laplacian.
The dynamics will tend to smooth firing
fronts, since concave regions of a firing front have more firing neurons around
them and will propagate faster than convex regions of the front.

We will restrict our discussion to the
case of the effectively random graph,
where the firing fraction $\Phi(t)$ is homogeneous in space.
The mean-field equation in this case is
\be \label{dynstep}
\frac{d\Phi}{dt} = (1 - \Phi)\theta \left(\Phi - \frac{m}{\kbar }\right)
\ee
In other words, below $m/\kbar$, $d\Phi/dt=0$ and the graph will never ignite, and above this value $%
d\Phi/dt \simeq 1$, which means that the graph fires within a few time steps
with the trivial exponential saturation (and that the continuous time
approximation is probably not very good, since the time scale is one step).

It is interesting to look more closely at the collectivity
function $\Psi(m,\Phi)$ of (\ref{Psi}), which was approximated in (\ref{dynstep}) by the
step-function. If $\kbar\gg 1$,
one can approximate the binomial distribution in (\ref{Psi}) by a normal one,
$N(k\Phi,\sigma)$, with a mean $k\Phi$ and a variance
$\sigma^2=k\Phi(1-\Phi)$. The
summation is replaced by integration,
\[
\sum_{\ell =m}^{k } \dbinom{k}{\ell}\Phi^{\ell}\ (1-\Phi )^{k-\ell}
=\frac{1}{\sigma\sqrt{2\pi}}\int_{m}^{\infty}\dr\ell\,
\exp[-(\ell -k\Phi)^2/2\sigma^2] ~.
\]
When all nodes have in-degree about $k$, that is, $p_k \simeq \delta_{k,\bar{%
k}}$, the collectivity function is approximated as in (\ref{slambda})
by the value of the integral at the boundary and we get
\[
\Psi(m,\Phi)\simeq \frac{1}{w\sqrt{2\pi}}\exp(-\HALF w^2)~, \quad w =\frac{m-\kbar\Phi}{\bigl(\kbar%
\Phi(1-\Phi)\bigr)^{1/2}}~.
\]
The dynamics is then
\be
\frac{d\Phi}{dt} = (1 - \Phi)\Psi(m,\Phi)=\HALF
(1-\Phi)\frac{1}{w\sqrt{2\pi}}\exp(-\HALF w^2)~.
\label{dynerf}
\ee

Taking into account in (\ref{dynerf}) fluctuations
in the number of firing inputs around the average value of
$k \Phi$ allows the possibility that a random graph will eventually ignite
even if the initial firing fraction $f = \Phi(t=0)$,
is much smaller than $\fs = m/\kbar$.
Of course, this ignition is a rather slow process.
In this regime, the dynamics (\ref{dynerf}) is approximately
\[
\frac{d\Phi}{dt} \simeq \left(\frac{\kbar \Phi}{m^2}\right)^{1/2}
                                      \exp\left(-{m^2}/({2\kbar\Phi})\right).
 \]
Neglecting the square root and integrating,
we find that the contribution is dominated by the initial value.
Thus, $\Phi$ is approximately
\[
\Phi(t) \sim \frac{m^2/(2\bar k)}{\log(\ts - t)}~,
\]
with $\ts \sim \exp \left({m^2}/(2\kbar f)\right)$.
Note that $\ts$ is the ``ignition time", \ie, the time to ignite a
given fraction of the graph, and it diverges exponentially as $f\to0$.

If $p_k$ is not sharp then one can utilize
the mean-field approximation and get
$\Psi(m,\Phi) = \int_{m}^{\infty } \dr k\, p_{k} \theta (k-m/\Phi)
=\int_{m/\Phi}^{\infty } \dr k\, p_{k}$,
which is simple to calculate for certain distributions.
For example, if $p_k$ distributes exponentially, $p_k \sim \exp(-k/\kbar)$,
then for small $\Phi \ll \fsran = m/\kbar$ the dynamics is approximately
$d\Phi/dt \simeq  \Psi \sim \exp(-{m}/{(\Phi\kbar)})$. The asymptotic growth is again logarithmic
$\Phi(t) \sim (m/\bar{k})/\log(\ts - t)$, with a somewhat different ignition time
$\ts \sim \exp\left({m}/({\kbar f})\right)$.

The asymptotic growth is quite different for scale-free in-degree distributions,
 $p_k \sim (k/k_0)^{-\gamma}$, where $k_0$ is a lower cut-off and $\gamma > 1$.
 In this case, the dynamics far from $\fs$ is
 $d\Phi/dt \sim (k_0\Phi/m)^{\gamma-1}$.
 Integrating,  one finds that, with initial condition $\Phi(0)=f$,
the growth in this case is a power law,
  \[
  \Phi(t) \sim \left(f^{2-\gamma} -\text{sgn}(\gamma-2)\cdot t/\tau \right)^{1/(2-\gamma)}~,
 \]
 with the positive constant $\tau = (m/k_0)^{\gamma-1}/|\gamma-2|$.
 The ignition time $\ts$ can be defined by $\Phi(\ts) = \fs$,
 which yields
 $$
\ts = \text{sgn}(\gamma-2)\cdot \tau (f ^{2-\gamma}-\fs^{2-\gamma}) ~.$$
If $1<\gamma<2$, this leads to
$$
\ts \sim  \tau \fs^{2-\gamma}~,\quad\text{and}\quad \Phi(t)\sim
(f^{2-\gamma}+ t/\tau)^{{1}/({{2-\gamma}})}~.
$$

For $\gamma>2$ the growth is slower and we get
 \[
 \ts\sim \tau / f^{\gamma-2}~,\quad\text{and}\quad\Phi(t)\sim
(1/f^{\gamma-2}- t/\tau)^{-{1}/({{\gamma-2}})}~.
 \]
In this case, the ignition time diverges as $f\to 0$.
 In the marginal case, $\gamma = 2$,
 we finally find
 \[
  \ts \sim (m/k_0) \log(\fs/f)~,\quad\text{and}\quad \Phi(t) \sim \exp\left({mt}/{k_0} \right)~.
  \]

It is interesting to note that exponential growth
was observed in experiments \cite{Eytan-2006},
which also suggested that the degree distribution is a power law.
However, it is not clear whether the simple model presented here can describe the dynamics
and whether the exponential growth is indeed related to a scale-free degree distribution.
It is evident from the last three examples that
the asymptotic growth below $\fs$ is determined by the tail
of the in-degree distribution.
Fat tail distributions lead to faster growth of the firing region
because they have a non-negligible fraction of ``hubs"
with many input edges.

\acknowledgements{We thank E. Moses for many insightful discussions.
This work was partially supported by Fonds National Suisse,
by the Israel Science Foundation, the Minerva Foundation, and the Clore Center.}

\bibliographystyle{apsrev}
\bibliography{QP}

\end{document}